\documentstyle[12pt]{article}

\def\1{\mbox{I\hspace{-.15em}1}}

\def\R{{\rm I\hspace{-.15em}R}}

\def\C{\hspace{3pt}{\rm l\hspace{-.47em}C}}

\def\b{\begin{equation}}
\def\e{\end{equation}}
\def\bee{\begin{enumerate}}
\def\eee{\end{enumerate}}
\title{Discrete Symmetries for Spinor Field\\in de Sitter
Space}
\author{S. Moradi$^1$,
S. Rouhani$^{2}$ and M.V. Takook$^{1,2}$ \thanks{e-mail:
takook@razi.ac.ir} }
\date{\today}

\begin{document}
\maketitle {\it \centerline{$^1$  Department of Physics, Razi
University, Kermanshah, IRAN}  \centerline{\it $^2$ Plasma Physics
Research Center, Islamic Azad University,}
 \centerline{\it P.O.BOX 14835-157, Tehran,
IRAN} }

\begin{abstract}

Discrete symmetries, parity, time reversal, antipodal, and charge
conjugation transformations for spinor field in de Sitter space,
are presented in the ambient space notation, {\it i.e.} in a
coordinate independent way. The PT and PCT transformations are
also discussed in this notation. The five-current density is
studied and their transformation under the discrete symmetries is
discussed.

\end{abstract}

\vspace{0.5cm} {\it Proposed PACS numbers}: 04.62.+v, 98.80.Cq,
12.10.Dm \vspace{0.5cm} 

\section{Introduction}

The interest in the de Sitter space increased when it turned out
that it could play a central role in the inflationary cosmological
paradigm \cite{li}. Very recently, the existence of a non-zero
cosmological constant has been proposed to explain the luminosity
observations on the farthest supernovas \cite{pe}. If this
hypothesis is validated in the future, our ideas on the
large-scale universe needs to be changed and the de Sitter metric
plays a further important role .

All these developments make it more compelling than ever to find a
formulation of de Sitter quantum field theory with the same level
of completeness and rigor as for its Minkowskian counterpart. In
Minkowski space, a unique Poincar\`e invariant vacuum can be fixed
by imposing the positive energy condition. In curved space-time,
however, a global time-like Killing vector field does not exist
and therefore the positive energy condition cannot be imposed.
Thus symmetry alone is not sufficient for determination of a
suitable vacuum state. In de Sitter space, however, symmetries
identify the vacuum only in relation to a two parameter ambiguity
$|\alpha,\beta\rangle$, corresponding to family of distinct de
Sitter invariant vacuum states for scalar field (see \cite{eila}
and references there in) . Only the one parameter family
$|\alpha,0\rangle$, is invariant under the PCT transformation
\cite{al,golo,spvo,bomast}. By imposing the condition that in the
null curvature limit the Wightman two-point function becomes
exactly the same as Minkowskian Wightman two-point function, the
other parameter $(\alpha)$ can be fixed as well. This vacuum,
$|0,0\rangle$, is called Euclidean vacuum or Bunch-Davies vacuum.
It should be noticed that this condition is different with the
Hadamard condition, which requires that the leading short distance
singularity in the Hadamard function $G^{(1)}$ should take its
flat space value. The leading singularity of the Hadamard function
is $\cosh 2\alpha$ times its flat space value \cite{eila2}.

Bros et al. \cite{brgamo,brmo} presented a QFT of scalar free
field in de Sitter space that closely mimics QFT in Minkowski
space. They introduced a new version of the Fourier-Bros
transformation on the one-sheeted hyperboloid \cite{brmo2}, which
allows us to completely characterize the Hilbert space of
``one-particle'' states and the corresponding irreducible unitary
representations of the de Sitter group. In this construction the
correlation functions are boundary values of analytical functions.

We generalized the Bros construction to the quantization of the
spinor free field in de Sitter space \cite{ta,ta1,brgamota}. In
the case of the interaction fields, the tree-level scattering
amplitudes of the scalar field, with one graviton exchange, has
been calculated in de Sitter space \cite{rota}. We have shown that
the $U(1)$ gauge theory in $4$-dimensional de Sitter space
describe the interaction between the spinor field (``electron'')
and the massless vector fields (``photon'') \cite{rota3}. In this
paper the discrete symmetries, parity, charge conjugation, time
reversal and antipodal transformation, for spionr field are
studied in the ambient space notation {\it i.e.} in a coordinate
independent way. In the next section, first we briefly recall the
appropriate notation. In section $3$, the discrete symmetries,
parity, time reversal, antipodal and PT transformations are
presented. In section $4$, charge conjugation transformations for
spinor field are discussed. The five-current density and Gordon
decomposition are calculated in section $5$. The transformation of
the five-current density under the discrete symmetries has been
discussed in this section as well. Finally, a brief conclusion and
outlook is set forth in section $6$. In the appendix, the discrete
symmetries in intrinsic global coordinates and their relation to
the ambient space notation are presented.

\section{Notation}

The de Sitter space is an elementary solution of the positive
cosmological Einstein equation in vacuum. It is conveniently seen
as a hyperboloid embedded in a five-dimensional Minkowski space
     \b X_H=\{x \in \R^5 ;x^2=\eta_{\alpha\beta} x^\alpha x^\beta
=-H^{-2}\},\;\;
      \alpha,\beta=0,1,2,3,4, \e
where $\eta_{\alpha\beta}=$diag$(1,-1,-1,-1,-1)$. The kinematical
group of the de Sitter space is the $10$-parameter group
$SO_0(1,4)$ and its contraction limit, $H=0$, is the Poincar\'e
group.

The spinor wave equation in de Sitter space-time has been
originally deduced by Dirac in 1935 \cite{dir}, and can be
obtained by the eigenvalue equation for the second order Casimir
operator \cite{ta,ta1,brgamota}
\begin{equation}  (-i\not x\gamma.\bar{\partial} +2i+\nu)\psi(x)=0,
\end{equation}
where $\not x=\eta_{\alpha \beta} \gamma^\alpha x^\beta$ and
$\bar{\partial_{\alpha}}=\partial_{\alpha}+H^2x_{\alpha}x\cdot\partial$.
In this notation we need the five  $4\times 4$ matrices
$\gamma^{\alpha}$, which are the generators of the Clifford
 algebra based on the metric $\eta_{\alpha\beta}$:
$$
\gamma^{\alpha}\gamma^{\beta}+\gamma^{\beta}\gamma^{\alpha}=2\eta^{\alpha\beta}{\1}\,
,
\quad{\gamma^{\alpha}}^{\dag}=\gamma^{0}\gamma^{\alpha}\gamma^{0}.\label{Clifford}
$$
An explicit quaternion representation, which is suitable for
symmetry consideration, is provided by \cite{ta,ta1}
$$ \gamma^0=\left( \begin{array}{clcr} \1 & \;\;0 \\ 0 &-\1 \\ \end{array} \right)
 ,\gamma^4=\left( \begin{array}{clcr} 0 & \1 \\ -\1 &0 \\ \end{array} \right)  $$
  \b   \gamma^1=\left( \begin{array}{clcr} 0 & i\sigma^1 \\ i\sigma^1 &0 \\    \end{array} \right)
     ,\gamma^2=\left( \begin{array}{clcr} 0 & -i\sigma^2 \\ -i\sigma^2 &0 \\  \end{array} \right)
      , \gamma^3=\left( \begin{array}{clcr} 0 & i\sigma^3 \\ i\sigma^3 &0 \\   \end{array} \right)\e
in terms of the $ 2\times2 $ unit $ \1 $ and Pauli matrices
$\sigma^i $. Due to the de Sitter group covariance, the adjoint
spinor is defined as follows \cite{ta1,brgamota,taka}:
\begin{equation} {\overline \psi}(x)\equiv
\psi^{\dag}(x){\gamma^0}{\gamma^4}.\label{adj}
\end{equation} dS-Dirac plane waves solutions are
\cite{ta,ta1,brgamota}  $$ \psi_1^{\xi,{\cal V}}(x)=(Hx.\xi )^{-2+
i \nu}{\cal V}(x,\xi)\equiv  (Hx.\xi )^{-2+ i \nu} H\not x\not
{\xi}{\cal V}^\lambda(\xi),$$
$$ \psi_2^{\xi,{\cal U}}(x)=(Hx.\xi)^{-2- i \nu}{\cal U}^\lambda(\xi),$$
where ${\cal V}^\lambda$ and ${\cal U}^\lambda$ are the
polarization spinors and
$$\xi \in {\cal C}^+=\{ \xi \;\;;\eta_{\alpha \beta}\xi^\alpha
\xi^\beta=(\xi^0)^2-\vec \xi.\vec \xi-(\xi^4)^2=0,\; \xi^0>0 \}.
$$ Due to the singularity and the sign ambiguity in phase value, the
solutions are defined in the complex de Sitter space
\cite{brgamo}, $$z \in X_H^{(c)}\equiv \{ z=x+iy\in \C^5
;\;\;\eta_{\alpha \beta}z^\alpha z^\beta=(z^0)^2-\vec z.\vec
z-(z^4)^2=-H^{-2}\}, $$ $$ \psi_1^{\xi,{\cal V}}(z)=(Hz.\xi )^{-2+
i \nu}{\cal V}(z,\xi),$$
$$ \psi_2^{\xi,{\cal U}}(z)=(Hz.\xi)^{-2- i \nu}{\cal U}(\xi).$$
The spinor field operator is defined by the boundary value of
complex solutions \begin{equation} \psi(z)=\int_T
\sum_{\lambda=1,2} \{\; a_\lambda(\xi,\nu){\cal
U}^\lambda(\xi)(z.\xi)^{-2-i\nu} +d^{\dag}_\lambda(\xi,\nu)H\not
z\not {\xi}{\cal V}^\lambda(\xi)(z.\xi)^{-2+i\nu}\; \}
d\mu_T(\xi),\e  where $T$ denotes an orbital basis of ${\cal C}^+$
and $ d\mu_T(\xi)$ is an invariant measure on ${\cal C}^+$
\cite{brmo}. The vacuum state, which is fixed by imposing the
condition that in the null curvature limit the Wightman two-point
function becomes exactly the same as Minkowskian Wightman
two-point function, is defined as follows \cite{brgamota}
$$a_\lambda(\xi,\nu)|\Omega>=0=d_\lambda(\xi,\nu)|\Omega>.$$ This vacuum, $|\Omega>$, is
equivalent to the Euclidean  vacuum $|0,0\rangle$. ``One particle
'' and ``one anti-particle'' states are \b
d^{\dag}_\lambda(\xi,\nu)|\Omega>=|\xi,\lambda,\nu>,\;\;\;a^{\dag}_\lambda(\xi,\nu)|\Omega>=
\overline{|\xi,\lambda,\nu>}.\e

\section{Discrete symmetries}

Many authors have thoroughly discussed the Kinematic invariance of
de Sitter space (de Sitter group), which explains the evolution of
various spin free fields \cite{dix,taka,moy,bodu}. The internal
symmetry, which describes the interaction between the spinor field
and the gauge boson field, has been studied recently \cite{rota3}.
Discrete symmetries for scalar field have been considered in
\cite{bodu} as well. At this stage we consider the discrete
symmetries, parity, time reversal and antipodal transformation,
for spionr field. The charge conjugation symmetry will be
considered in the next section.

In accordance with the invariance principle, we shall verify that
the form of dS-Dirac field equation remains unaltered in two
frames, related by the above transformations. Let our system be
described by the wave function $\psi$ in the first frame and by
$\psi'$ in the transformed frame. Both must satisfy the dS-Dirac
equation, \b (-i\not x'\gamma.\bar{\partial'} +2i+\nu)\psi'(x')=0,
\;\; x'=\Lambda x.\e There must be a local relation between $\psi$
and $\psi'$, so that the observer in the second frame may
reconstruct $\psi'$ by $\psi$. We assume that this relation is
linear: \b \psi'(x')=S(\Lambda)\psi(x),\e where $S(\Lambda)$ is a
nonsingular $4 \times 4$ matrix. Equation $(7)$ now reads \b (-
S^{-1}(\Lambda)i\not x'\gamma.\bar{\partial'}S(\Lambda)
+2i+\nu)\psi(x)=0.\e If this equation is be equivalent to $(2)$,
for any $\psi$, we must have \b S^{-1}(\Lambda)  \not
x'\gamma.\bar{\partial'} S(\Lambda)= \not
x\gamma.\bar{\partial}.\e Construction of $S(\Lambda)$ will have
to comply with the following transformations.

\subsection{Parity }

Space reflection or parity transformation, in de Sitter ambient
space, is defied by the following relation \cite{bodu}
$$x=(x^0, \vec x, x^4) \longrightarrow x_p=(x^0, -\vec x, x^4)\equiv \Lambda_p x$$
where the matrix $\Lambda_p$ is
$$\Lambda_p=\left( \begin{array}{clccr} 1 & 0& 0& 0&0 \\
0 & -1& 0& 0&0 \\ 0 & 0& -1& 0&0\\ 0 & 0& 0& -1 &0\\0 & 0& 0& 0& 1
\\ \end{array} \right).$$ Since $S(\Lambda_p)$
should satisfy $(10)$, by use of the relation $\not x_p=\gamma^0
\gamma^4 \not x \gamma^4 \gamma^0$, it is easily verified that the
simple choice is the desired transformation \b
\psi'(x_p)=S(\Lambda_p)\psi(x)=\eta_p \gamma^0 \gamma^4 \psi(x),\e
where $\eta_p$ is an arbitrary, unobservable phase quantity. Since
in the null curvature limit, we have \cite{ta,brgamota}
\b(\gamma^\mu \gamma^4)_{dS}\equiv(\gamma^\mu)_{M}; \;\;
\mu=0,1,2,3,\e the parity is exactly the same as in the
Minkowskian space. Therefore through the parity transformation in
classical spinor field, we have \b S(\Lambda_p)=\eta_p \gamma^0
\gamma^4.\e  From the quantum field point of view, one can find an
operator $U_p$ which satisfies the following:
$$ U_p\psi(x)U_p^\dag=S(\Lambda_p)\psi( x_p).$$

\subsection{Time reversal}

The time reversal transformation in de Sitter ambient space is
defied by the following relation \cite{bodu}
$$x=(x^0, \vec x, x^4) \longrightarrow x_t=(-x^0, \vec x, x^4).$$
It could be shown by
$$\Lambda_t=\left( \begin{array}{clccr} -1 & 0& 0& 0&0 \\
                                         0 & 1& 0& 0&0 \\
                                         0 & 0& 1& 0&0\\
                                        0 & 0& 0& 1 &0\\
                                        0 & 0& 0& 0& 1
 \\ \end{array} \right).$$  $S(\Lambda_t)$ should
satisfy $(10)$. By the use of the relation $\not x_t=- \gamma^0
\not x \gamma^0$, it is easy to see that the simple choice is \b
S(\Lambda_t)=\eta_t \gamma^0,\e where $\eta_t$ is an arbitrary,
unobservable phase value. Therefore $S(\Lambda_t)$ must be the
time reversal transformation for the classical spinor field. From
the quantum field point of view, one can find an operator $U_t$
satisfying $$ U_t\psi(x)U_t^\dag=S(\Lambda_t)\psi( x_t).$$  $U_t$
is antiunitary and antilinear operator, in the Minkowski space,
which preserves the sign of energy under the time reversal
transformation. In de Sitter space, however,  the concept of
energy is not defined globally and one can not impose the positive
energy condition for obtaining an antiunitary and antilinear
operator $U_t$.

\subsection{PT transformation}

The PT transformation in de Sitter ambient space is defied by the
following relation \cite{bodu}
$$x=(x^0, \vec x, x^4) \longrightarrow x_{pt}=(-x^0, -\vec x, x^4)\equiv \Lambda_{pt} x$$
where the matrix $\Lambda_{pt}$ is
$$\Lambda_{pt}=\left( \begin{array}{clccr}- 1 & 0& 0& 0&0 \\
0 & -1& 0& 0&0 \\ 0 & 0& -1& 0&0\\ 0 & 0& 0& -1 &0\\0 & 0& 0& 0& 1
\\ \end{array} \right).$$ $S(\Lambda_{pt})$
should satisfy $(10)$. By the use of the relation $\not
x_{pt}=-\gamma^4 \not x \gamma^4 $, it is easy to see that \b
\psi'(x_{pt})=S(\Lambda_{pt})\psi(x)=\eta_{pt} i\gamma^4 \psi(x)\e
is the desired transformation. Here $\eta_{pt}$ is an arbitrary,
unobservable phase value.

\subsection{Antipodal transformation}

The Antipodal transformation in de Sitter ambient space is defied
by the following relation \cite{bodu}
$$x=(x^0, \vec x, x^4) \longrightarrow x_a=(-x^0, -\vec x, -x^4).$$
It is presented by the matrix $\Lambda_a$,
$$\Lambda_a=\left( \begin{array}{clccr} -1 & 0& 0& 0&0 \\
                                         0 & -1& 0& 0&0 \\
                                         0 & 0& -1& 0&0\\
                                        0 & 0& 0& -1 &0\\
                                        0 & 0& 0& 0& -1
 \\ \end{array} \right)=-I,$$ where $I$ is the unite $4 \times 4$ matrix. $S(\Lambda_a)$
should satisfy $(10)$. By the use of the relation $\not x_a=- \not
x $, it is easy to see that \b
\psi'(x_a)=S(\Lambda_a)\psi(x)=\eta_a I \psi(x)\e is the desired
transformation. Here $\eta_a$ is an arbitrary, unobservable phase
value. Therefore $S(\Lambda_a)$ is the antipodal transformation
for the classical spinor field, \b S(\Lambda_a)= \eta_a I\equiv
(-1)^nI.\e It clearly seen that $\psi(x)$ is a homogeneous
function of $\R^5$-variables $x^{\alpha}$, with homogeneity degree
$n$ \cite{dir}. This transformation has no analogy in Minkowski
space.

The full de Sitter group has four connected component: $SO_0(1,4),
\Lambda_p.SO_0(1,4), \Lambda_t.SO_0(1,4)$ and $
\Lambda_{pt}.SO_0(1,4)$. Since $ \Lambda_t.SO_0(1,4)\equiv
\Lambda_a.SO_0(1,4)$, due to the fact that $\Lambda_a.\Lambda_t\in
SO_0(1,4)$, $\Lambda_a $ can be obtained by some rotations of
angle $\pi$ of the spacelike coordinate $ \vec x, x^4$. Theses two
discrete transformations $\Lambda_a$ and $\Lambda_t$ are closely
linked together. $\Lambda_t$ violation in de Sitter invariant
theories will imply violation of $\Lambda_a$. Thus the
CP-violating effects in K-meson decay would bring a drastic
additional symmetry breaking namely either violation of both
$\Lambda_t$ and $\Lambda_a$ (where CPT is conserved,) or
$\Lambda_t$ is conserved and consequently CP and CPT are violated
\cite{bodu}. The axiomatic approach to the scalar field
quantization in de Sitter space and their consequences to the CPT
theorem is discussed by Bros et al \cite{brmo,brmo2}.

\section{Charge conjugation}

The $U(1)$ local gauge invariant spinor field equation is obtained
in a coordinate independent way notation \cite{rota3} {\it i.e.}
\b (-i\not x \not \bar{\partial} - q\not x \not K(x)+2i+\nu
)\psi(x)= 0,\e where $K(x)$ is a ``massless'' vector field
(electromagnetic field) \cite{gata,gagata1,garota} and $q$ is a
free parameter which can be identified with the electric charge in
the null curvature limit. In Minkowski space, the Dirac equation
has a symmetry corresponding to the ``particle $\leftrightarrow$
antiparticle''interchange. This transformation is known as the
charge conjugation. In de Sitter space the concept of particle,
antiparticle, and charge are not defined clearly.

Similar to the Minkowski space, we have defined the charge
conjugation transformation for de Sitter ambient space by the
following relation $ \left\{\begin{array}{clcr} q &
\longrightarrow -q \; \\ \nu & \rightarrow -\nu\\ \end{array}
\right..$ From the point of view of the representation theory of
de Sitter group, the representation $U^{\nu,s}$ and $U^{-\nu,s}$
are equivalent \cite{dix,taka,moy,gata}. There is no difference
between classical spinor fields with opposite sign of $\nu$. But
in the quantum field representation $(5) $, one can see that the
opposite sign of the $\nu$ corresponds to the particle and
antiparticle interchange $(6)$. In other words quantized fields
may describe particles of opposite charge with identical spin and
opposite sign of $\nu$.

We thus seek a transformation $\psi \rightarrow \psi^c$, reversing
the ``charge'' such that \b (-i\not x\not {\bar{\partial}} +q\not
x\not K+2i-\nu)\psi^{c}(x)=0.\e We demand that this transformation
should be local. To construct $\psi^c$ we conjugate and transpose
the equation $(18)$ and get $$
 \bar \psi\gamma^4(i
\overleftarrow{\not\bar\partial} \not x-q\not K\not x-2i+\nu)=0
,$$ \b (-i (\not x)^T(\not\bar\partial)^T +q(\not x)^T(\not
K)^T+2i- \nu)(\gamma^4)^T(\bar \psi)^T=0 ,\e where $(\bar
\psi)^T=(\gamma^4)^T(\gamma^0)^T \psi^*$. In any representation of
the $\gamma$ algebra there must exist a matrix $C$ which satisfies
\b
C(\gamma^\alpha)^T(\gamma^\beta)^TC^{-1}=\gamma^\alpha\gamma^\beta,
\;\; \mbox{or} \;\;C\gamma_\alpha^T C^{-1}=\pm\gamma_\alpha.\e In
order to obtain the Minkowskian charge conjugation in the null
curvature limit, the negative sign is chosen.  We then identify
$\psi^c$ as \b \psi^{c}=\eta_c C(\gamma^4)^T(\bar \psi)^T,\e where
$\eta_c$ is an arbitrary unobservable phase value, generally
chosen to be unity. In the present framework charge conjugation is
an antilinear transformation. In the $\gamma$ representation $(3)$
we have: $$ C\gamma^{0}C^{-1}=-\gamma^0 ,
C\gamma^4C^{-1}=-\gamma^4$$ \b
C\gamma^{1}C^{-1}=-\gamma^1,C\gamma^{3}C^{-1}=-\gamma^3,C\gamma^{2}C^{-1}=\gamma^2.
\e In this representation $C$ commutes with $\gamma^2$ and
anticommutes with other $\gamma$-matrix therefore the simple
choice may be taken as $C=\gamma^2$. It satisfies \b
C=-C^{-1}=-C^T=-C^\dag.\e This clearly illustrates the non
singularity of $C$. From the quantum field point of view one can
fined an operator $U_c$ satisfying
$$ U_c\psi U_c^\dag=\psi^c=\eta_c C(\gamma^4)^T(\bar
\psi)^T.$$

The adjoint spinor, which is defined by ${\overline \psi}(x)\equiv
\psi^{\dag}(x){\gamma^0}{\gamma^4}$, transforms in a different way
from $\psi$, under de Sitter transformation \cite{brgamota} $$
\psi'(x')=S(\Lambda)\psi(x),\;\; \bar \psi'(x')=-\bar
\psi(x)\gamma^4 S^{-1}(\Lambda)\gamma^4,$$ \b x'=\Lambda x,\;\;
\Lambda \in SO(1,4), \;\; S \in Sp(2,2).\e On the contrary it is
easy to show that $\psi^c$ transforms in the same way as $\psi$
$$\psi'^c(x')=S(\Lambda)\psi^c(x),$$ which is an important result.
According to $(19)$, the wave equation of $\psi^c$, is different
from the wave equation of $\psi$ by the signs of the $q$ and
$\nu$. Consequently if $\psi$ describes the motion of a dS-Dirac
"particle'' with the charge $q$, $\psi^c$ represents the motion of
a dS-Dirac "anti-particle'' with the charge $(-q)$. In other
words, $\psi$ and ${\psi}^c$ can described as particle and
antiparticle functions. A very interesting result of this
representation is that $\psi$ and ${\psi}^c$ are charge
conjugation of each other \b ({\psi}^c)^c=\gamma^0
C{{\psi}^c}^\ast=\psi. \e

A fundamental property of local quantum theory in Minkowski space,
proved by Pauli, Zumino and Schwinger, states that in all cases,
the action is invariant under $\Theta\equiv U_pU_cU_t$
transformation. This is the famous PCT conservation theorem. The
PCT transformation in de Sitter ambient space is defied by the
following relation  \b \left\{\begin{array}{clcr}
q   \\\nu\\ x=(x^0, \vec x, x^4)  \\
\end{array} \right.  \longrightarrow \left\{\begin{array}{clcr}
-q  \\ -\nu \\x'=(-x^0, -\vec x, x^4).  \\
\end{array} \right. \e The construction of the operator $\Theta$
and the transformation of the two parameter spinor-vacuum
$|\alpha,\beta\rangle$ and the quantum spinor field will be
considered in the forthcoming paper.

\section{The five-current density}

In order to discuss dS-interaction field theory in the ambient
space notation, we need to know the five-current density and its
different transformations under different symmetries. Recently we
have obtained the five-current density by using the Noether's
theorem \cite{rota3}. Alternatively we have obtained here, the
five-current density by using the dS-Dirac equation and its
adjoint. The dS-Dirac equation for $ \bar\psi$ is \cite{brgamota}:
\b
 \bar \psi\gamma^4(i
\overleftarrow{\not\bar\partial} \not x-2i+\nu)=0. \e Combining
Eqs. $(2)$ and $(28)$ and using $\bar \psi \gamma^4 \psi=0$
(\cite{brgamota} Appendix C), lead to: \b
\bar{\partial_\alpha}(\bar\psi\gamma^4\gamma^\alpha\not x\psi)=0
\e We have, therefor, a candidate for the current \b
J^\alpha=H\bar\psi\gamma^4\gamma^\alpha\not
x\psi\equiv-H\bar\psi\gamma^4\not x \gamma^\alpha \psi ,\e which
is exactly the same as the conserved current \cite{rota3}. It can
be easily shown that $ x\cdot J =H\bar\psi\gamma^4\psi =0$.

We now consider the transformation of the five-current density
under different symmetries. The current density, $J^\alpha$, is
transformed as a five vector under de Sitter transformation: \b
J'^\alpha=H{\bar\psi}'\gamma^4\gamma^\alpha{\not x}'{\psi}' =
H\bar \psi(-\gamma^4S^{-1}\gamma^4)\gamma^4\gamma^\alpha S\not x
S^{-1} S\psi=\Lambda^\alpha_\beta J^\beta, \e where we have used
the relation $S^{-1}\gamma^\alpha S =\Lambda^\alpha_\beta
\gamma^\beta$ \cite{brgamota}. Therefore the current density is a
vector field on the de Sitter hyperboloid. Its transformation
under the charge conjugation is:\b
{J_c}^\alpha=H{\bar{\psi}}^c\gamma^4\gamma^\alpha\not
x\psi^c=H\bar\psi\gamma^4\gamma^\alpha\not x \psi =J^\alpha, \e
where $\psi^c = \gamma^0 C\psi^\ast$ and ${\bar{\psi}}^c=\psi^T
\gamma^4 C$. Under the time reversal, the transformation is \b
J_t^\alpha(x_t)=H \psi'^\dag\gamma^4\gamma^\alpha\not
x_t\psi'=(\Lambda_t) ^\alpha_\beta J^\beta(x). \e Under the parity
transformation, we have \b J_p^\alpha(x_p)=
H\psi'^\dag\gamma^4\gamma^\alpha\not x_p\psi'=-(\Lambda_p)
^\alpha_\beta J^\beta(x), \e where $J_p^\alpha$ is a
pseudo-vector.

It is easy to show that $\phi(x)=\psi^\dag\gamma^0\not
x\psi=-\bar\psi\gamma^4 \not x\psi$ is a scalar field under de
Sitter transformation \b \bar\psi'\gamma^4 \not x'\psi'=-\bar \psi
\gamma^4 S^{-1}\gamma^4\gamma^4\not x'S\psi=\bar \psi \gamma^4
S^{-1}\not x'S\psi=\bar\psi\gamma^4 \not x\psi. \e The
transformation under parity is \b
\phi'(x_p)=\psi'^\dag\gamma^0\not
x_p\psi'=\psi^\dag\gamma^0\gamma^4\gamma^0\not
x_p\gamma^0\gamma^4\psi=\psi^\dag\gamma^0\not x\psi.\e The
behavior of the scalar density under the charge conjugation is \b
\phi^c={\psi^c}^\dag \gamma^0\not x\psi^c= \psi^{\ast T}\not
x^{\ast T}\gamma^{0T}\psi=\psi^{\dag}\not
x^{\dag}\gamma^0\psi=\psi^{\dag}\gamma^0\not x\psi\e

Finally it is interesting to calculate the dS-Gordon
decomposition. By multiplying the adjoint dS-Dirac equation from
right by $\not x\not a\psi$ and dS-Dirac equation from left by
$\bar \psi\gamma^4\not a\not x$, we obtain \b \bar
\psi\gamma^4(iH^{-2}\not a\not \bar\partial +2i\not a\not
x+\nu\not a\not x)\psi=0\e\b\bar \psi\gamma^4(iH^{-2}
\overleftarrow{\not\bar\partial}\not a +2i\not x\not a-\nu\not
x\not a )\psi=0.\e Combining these equations we obtain: \b iH^{-2}
\bar \psi\gamma^4(\not a\not
\bar\partial+\overleftarrow{\not\bar\partial}\not a)\psi +2\nu\bar
\psi\gamma^4\not a\not x\psi=0.\e By using $\not a\not b=a\cdot
b+2ia_\alpha b_\beta S^{\alpha\beta}$, where $
S_{\alpha\beta}=-{i\over
4}\lbrack\gamma_{\alpha},\gamma_{\beta}\rbrack\label{genspin} $ we
can write \b\not a\not
\bar\partial=a\cdot\bar\partial+2ia_\alpha\bar\partial _\beta
S^{\alpha\beta}\e\b\overleftarrow{\not\bar\partial}\not
a=a\cdot\overleftarrow{\bar\partial}-2ia_\alpha
\overleftarrow{\bar\partial}_\beta S^{\alpha\beta}.\e By
eliminating $a^\alpha$, we can writ the five-current density in
the following form \b J^\alpha=H\bar\psi\gamma^4\gamma^\alpha\not
x\psi=\frac{1}{H\nu}[\bar\psi\gamma^4
S^{\alpha\beta}({\bar\partial_\beta}\psi)-(\bar\partial_\beta{\bar\psi})\gamma^4
S^{\alpha\beta}\psi].\e

\section{Conclusion and outlook}

The formalism of the quantum field theory in de Sitter universe,
in ambient space notation, is very similar to the quantum field
formalism in Minkowski space. An explicit form of the charge
conjugation matrix $C$, associated with the universal covering
group of $SO_0(1,4)$, {\it i.e.} $Sp(2,2)$, has been obtained.
Equipped with the results of the present paper, various
interaction Lagrangians between the fermions and bosons,
preserving the de Sitter invariance and discrete symmetries can
easily be defined. Using charge conjugation matrix $C$ one can
study the supersymmetry algebra in de Sitter space as well. The
importance of this formalism may be shown further by the
consideration of the linear quantum gravity \cite{gagata2} and
supergravity in de Sitter space, which lays a firm ground for
further study of evolution of the universe.

\vskip 0.5 cm \noindent {\bf{Acknowledgements}}: The authors would
like to thank A. Pahlavan for interest in this work. They would
like to thank the referee for very useful suggestions.

\begin{appendix}

\section{Discrete symmetries in intrinsic global coordinate}

In de Sitter space one can define four type of coordinate systems:
global, flat, open and static coordinates \cite{bida}. It is the
global type coordinate that cover the whole de Sitter manifold.
Therefor, only in this type of coordinates, the discrete
symmetries can be defined properly \cite{ta}. If we chose the
global coordinates as:
$$ \left\{\begin{array}{clcr} x^0&=H^{-1}\sinh Ht \\
                                  x^1&=H^{-1}\cosh Ht\cos\chi\\
                         x^2&=H^{-1}\cosh Ht\sin\chi \cos\theta\\
                         x^3&=H^{-1}\cosh Ht\sin\chi \sin\theta\cos\phi \\
          x^4&=H^{-1}\cosh Ht\sin\chi\sin\theta\sin\phi\\
         \end{array} \right.$$
with the metric
       $$ ds^2=\eta_{\alpha\beta}dx^\alpha dx^\beta\mid_{x^2=-H^{-2}}=
       g_{\mu\nu}dX^\mu dX^\nu$$ $$ =dt^2-H^{-2}\cosh^2 Ht [d\chi^2+\sin^2 \chi (d \theta^2
              +\sin^2\theta d\phi^2)],$$
the discrete symmetries are defined by: $$\begin{tabular}{|c|c|c|}
  \hline
  antipodal & time reversal &  parity \\
  \hline
  $X^0=t \longrightarrow -t$ &$t \longrightarrow -t$ & $t \longrightarrow t$ \\
  $X^1=\chi \longrightarrow \pi-\chi$ &$\chi \longrightarrow \chi$ & $\chi \longrightarrow \pi-\chi$\\
  $X^2=\theta \longrightarrow \pi-\theta$ & $\theta \longrightarrow \theta$ & $\theta \longrightarrow \pi-\theta$ \\
  $X^3=\phi \longrightarrow \pi+\phi$ & $\phi \longrightarrow \phi$ & $\phi \longrightarrow \pi-\phi$ \\
  \hline
\end{tabular}$$
The Dirac equation in the general curved space time is \cite{na}
\b
(\gamma^{\mu}(X)\nabla_{\mu}-m)\Psi(X)=0=(\bar{\gamma}^{a}\nabla_{a}-m)\Psi(X),\e
where
$$\{\gamma^{\mu}(X),\gamma^{\nu}(X)\}=2g^{\mu\nu},\;\{\bar{\gamma}^{a},\bar{\gamma}^{b}\}=
2\eta^{ab} ,\;\; \mu,a=0,1,2,3.$$  Here $\nabla$ is the spinor
covariant derivative
$$ \nabla_{a}\Psi(X)={e^{\mu}}_{a}(\partial_{\mu}+
\frac{i}{2}{e^{c}}_{\nu}\nabla_{\mu}e^{b\nu}\Sigma_{cb})\Psi(X),$$
where ${e_{\mu}}^{a}$ is the local vierbein, ${e_{\mu}}^{a}
{e_{\nu}}^{b}\eta_{ab}=g_{\mu\nu}$, and $
\Sigma_{cb}=\frac{i}{4}[\bar{\gamma}_{c},\bar{\gamma}_{b}]$ is the
spinor representation of the generators of the Lorentz
transformation. The Dirac equation under the discrete symmetries
transform in the following form
$$(\gamma^{\mu}(X')\nabla'_{\mu}-m)\Psi'(X')=0=(\bar{\gamma}^{a}\nabla'_{a}-m)\Psi'(X').$$ By using the
linear relation $\Psi'(X')={\cal S}\Psi(X)$ we obtained the
condition $$ {\cal S}^{-1}\bar{\gamma}^{a} \nabla'_{a}{\cal
S}=\bar{\gamma}^{a}\nabla_{a}, \; \;\mbox{or}\;\; {\cal
S}^{-1}{\gamma}^{\mu}(X') \nabla'_{\mu}{\cal
S}={\gamma}^{\mu}(X)\nabla_{\mu}.$$ Through a direct calculation,
the discrete symmetries in this notation can be obtained:
$$
\begin{tabular}{|c|c|c|}
  \hline
  time reversal & parity & antipodal  \\
  \hline
  ${\cal S}(t)\bar{\gamma}^{0}{\cal S}(t)^{-1}=-\bar{\gamma}^{0}$ & ${\cal S}(p)\bar{\gamma}^{0}{\cal S}(p)^{-1}=\bar{\gamma}^{0}$ & ${\cal S}(a)\bar{\gamma}^{0}{\cal S}(a)^{-1}=-\bar{\gamma}^{0}$   \\
  ${\cal S}(t)\bar{\gamma}^{1}{\cal S}(t)^{-1}=\bar{\gamma}^{1}$ & ${\cal S}(p)\bar{\gamma}^{1}{\cal S}(p)^{-1}=-\bar{\gamma}^{1}$ & ${\cal S}(a)\bar{\gamma}^{1}{\cal S}(a)^{-1}=-\bar{\gamma}^{1}$  \\
 ${\cal S}(t)\bar{\gamma}^{2}{\cal S}(t)^{-1}=\bar{\gamma}^{2}$ & ${\cal S}(p)\bar{\gamma}^{2}{\cal S}(p)^{-1}=-\bar{\gamma}^{2}$ & ${\cal S}(a)\bar{\gamma}^{2}{\cal S}(a)^{-1}=-\bar{\gamma}^{2}$   \\
 ${\cal S}(t)\bar{\gamma}^{3}{\cal S}(t)^{-1}=\bar{\gamma}^{3}$ & ${\cal S}(p)\bar{\gamma}^{3}{\cal S}(p)^{-1}=-\bar{\gamma}^{3}$ & ${\cal S}(a)\bar{\gamma}^{3}{\cal S}(a)^{-1}=\bar{\gamma}^{3}$  \\
  $\Psi'(X_t)= \eta_t \bar \gamma^1\bar \gamma^2\bar \gamma^3 \Psi(X)$ & $\Psi'(X_p)= \eta_p \bar \gamma^0 \Psi(X)$ & $\Psi'(X_a)= \eta_a i\bar \gamma^3 \Psi(X)$  \\
  \hline
\end{tabular}
$$
where $\eta$'s are the arbitrary, unobservable phase values. In
the next appendix we discus the relation between the two
formalisms.

\section{The relation between the intrinsic and ambient space notation }

First we briefly recall here the relation between de Sitter-Dirac
equation ($2$) and the usual Dirac equation for curved spacetimes
obtained by the method of covariant derivative $(44)$, which is
established in the paper by G\"ursey and Lee \cite{gule}.

They introduced a set of coordinates $\{y^{\alpha}\}\equiv (y^\mu,
y^4=H^{-1})$ related to the $\{x^{\alpha}\}$'s by $$ x^{\alpha}=
\bigl( Hy^{4} \bigl) f^{\alpha}(y^{0},y^{1},y^{2},y^{3}), $$ where
arbitrary functions $f^{\alpha}$ satisfying
$f^{\alpha}f_{\alpha}=-H^{-2}$. Introducing five matrices $
\beta^{\alpha}\equiv \left( \frac{\partial y^{\alpha}}{\partial
x^{\beta}} \right) \gamma^{\beta},$ satisfy the anticommutation
properties $$ \{ \beta^{\mu},\beta^{\nu}\}=2g^{\mu \nu},\; \{
\beta^{\mu},\beta^{4}\}=0 $$ with $g^{\mu
\nu}=\eta^{\alpha\beta}\frac{\partial y^{\mu}}{\partial
x^{\alpha}}\frac{\partial y^{\nu}}{\partial x^{\beta}}$,
$\mu,\nu=0,\dots ,3$, and $\psi= (1\pm i\beta^{4})\chi, $   $ \chi
$ satisfies the G\"ursey--Lee equation
\begin{equation}
\left( \beta^{\mu} \frac{\partial}{\partial y^{\mu}} -2H\,
\beta^{4}-m \right) \chi =0, \label{GurLee}
\end{equation}
where $m=H\nu$. Choosing at every point of de Sitter spacetime, a
local vierbein ${e_{\mu}}^{a}$, and setting $\gamma^{\mu}(X)\equiv
{e^{\mu}}_{a}\gamma^{a}$, there exists a transformation $V$ such
that $ \gamma^{\mu}(X)=V\beta^{\mu}(y)V^{-1}. $ Then, under the
$V$ transformation, the G\"ursey--Lee equation $(45)$ becomes
equation $(44)$ with $\Psi(X)=V\chi(y).$ It is interesting to
notice that the matrix $\beta^4=\gamma_\alpha x^\alpha$ is related
to the constant matrix $\gamma^4$ by $\gamma^{4}=V\beta^{4}V^{-1}$
\cite{gule}.

Now we can write the relation between the spinor field in two
notations $$ \psi(x)=V^{-1}(1\pm i\gamma^4)\Psi(X).$$ It is
important to notice that the inverse of matrix $V^{-1}(1\pm
i\gamma^4)$ does not exist-similar to the relation between tensor
field in the two notations \cite{gata}. The matrix that transforms
the $\psi(x) $ to $\Psi(X)$ can be calculated as well. By using
the above equation and the transformations of the local vierbein
${e_{\mu}}^{a}$, the relation between the discrete symmetries in
the two formalisms can be obtained directly.
\end{appendix}

\end{document}